\pdfoutput=1
\documentclass[twocolumn,preprintnumbers,amsmath,amssymb,letter]{revtex4}
\usepackage{graphics}
\usepackage{graphicx}
\usepackage{amsmath}
\usepackage{amssymb}
\usepackage{bm}

\newcommand{\rar}{\rightarrow}
\newcommand{\ra}{\rangle}
\newcommand{\la}{\langle}

\newcommand{\mb}{\mathbb}

\newtheorem{proposition}{Proposition}
\newcommand{\qed}{\hfill $\Box$ \hfill \\}

\begin{document}

\preprint{LAUR-07-5287}

\title{Mapping Semantic Networks to Undirected Networks}

\author{Marko A. Rodriguez \\
		T-7, Center for Non-Linear Studies \\
		Los Alamos National Laboratory \\
		Los Alamos, New Mexico 87545}

 \begin{abstract}
There exists an injective, information-preserving function that maps a semantic network (i.e~a directed labeled network) to a directed network (i.e.~a directed unlabeled network). The edge label in the semantic network is represented as a topological feature of the directed network. Also, there exists an injective function that maps a directed network to an undirected network (i.e.~an undirected unlabeled network). The edge directionality in the directed network is represented as a topological feature of the undirected network. Through function composition, there exists an injective function that maps a semantic network to an undirected network. Thus, aside from space constraints, the semantic network construct does not have any modeling functionality that is not possible with either a directed or undirected network representation. Two proofs of this idea will be presented. The first is a proof of the aforementioned function composition concept. The second is a simpler proof involving an undirected binary encoding of a semantic network.
\end{abstract}

\maketitle{}

\section{Introduction}

A network is a popular data structure for representing the relationship between discrete elements \cite{netanal:brandes2005,netsci:newman2006}. There are various types of networks such as the undirected network (i.e.~undirected unlabeled network), the directed network (i.e.~directed unlabeled network), and the semantic network (i.e.~directed labeled network). In an undirected network, there exists no order to the relationships between the vertices. An undirected network can be denoted $U \subseteq \{V^u \times V^u\}$, where $V^u$ is the vertex set and any edge $\{i,j\} \in U$ denotes an undirected relationship. The directed network provides the concept of edge directionality. A directed network can be represented as $D \subseteq (V^d \times V^d)$, where $V^d$ is the vertex set and any edge $(i,j) \in D$ denotes a directed relationship. All edges in both an undirected and directed network are homogeneous in meaning. In order to represent edge meaning, a semantic network can be used. In a semantic network, an edge connecting any two vertices maintains a label (e.g.~character string) that denotes the type of relationship between two vertices. A semantic network can be represented as $S \subseteq \la V^s \times \Omega \times V^s \ra$, where $V^s$ is the vertex set, $\Omega$ is the set of edge labels, and any edge (called a triple) $\la i,\omega,j \ra \in S$ denotes an ordered, labeled relationship.

The semantic network is perhaps best known as a modeling construct from the early days of knowledge representation in the cognitive sciences \cite{sowa:semantic1991}. However, with the inception of the Semantic Web initiative \cite{lee:semantic2001,pubsem:lee2001} and with the development of triple-store technology (i.e.~semantic network databases) \cite{lee:triple2004,oracle:alexander2006,agraph:aasman2006}, there has been an increase in the use of the semantic network as a data structure for modeling data sets where there exists a heterogeneous set of vertices and edges. This trend has been occurring across various disparate domains such as bioinformatics \cite{sembio:quan2003,sembio:ruttenberg2007}, digital libraries \cite{lib:bax2004,semever:bollen2007}, and general computer-science \cite{rodriguez:gpsemnet2007}. Because of the use of the labeled edge, the semantic network is seen as the better modeling construct  than both the undirected and directed network for such data sets.

However, when ignoring space constraints, there is no modeling gain by using a semantic network representation as opposed to a directed network representation. Moreover, there is no modeling gain over using an undirected network representation. Through a series of information-preserving, injective mappings \footnote{An injective function is one such that if $f(a) = f(b)$, then $a = b$.}, this article demonstrates that it is possible to model a semantic network both as a directed and undirected network. While the directed and undirected models of a semantic network utilize more vertices and edges in their representation, they ultimately have the ability to capture the same information.

The outline of this article is as follows. Section \ref{sec:sem-to-dir} presents an injective function to map a semantic network to a directed network. Section \ref{sec:dir-to-und} presents an injective function to map a directed network to an undirected network. Finally, through function composition, Section \ref{sec:sem-to-und} presents an injective function to map a semantic network to an undirected network.

\section{Mapping a Semantic Network to a Directed Network\label{sec:sem-to-dir}}

This section will present an injective, information-preserving function that maps a semantic network to a directed network. There is a two step process to this function. First, the edge labels of a semantic network are represented as a binary string. Second, each binary string is represented as a unique directed network encoding. Given that a directed network can only represent vertices and directed edges, each edge label of the semantic network is encoded as a topological feature in the directed network.

\begin{figure}[h!]
	\centering
	\includegraphics[width=0.175\textwidth]{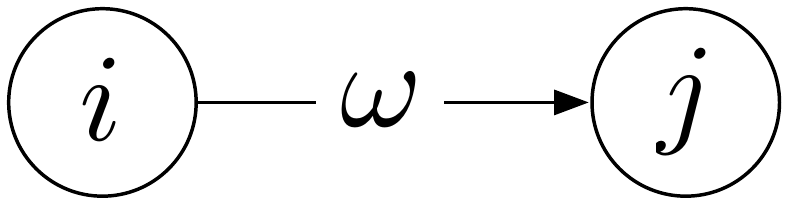}
	 \caption{\label{fig:semantic-example}An edge in a semantic network.}
\end{figure}

Let $S \subseteq \la V^s \times \Omega \times V^s \ra$ denote a semantic network where $V^s$ is the set of all vertices and $\Omega$ is the set of all edge labels. Any triple $\la i, \omega, j \ra \in S$ represents a directed edge from vertex $i$ to vertex $j$ with a label of $\omega$.  An example semantic network triple is diagrammed in Figure \ref{fig:semantic-example}.

\begin{figure}[h!]
	\centering
	\includegraphics[width=0.175\textwidth]{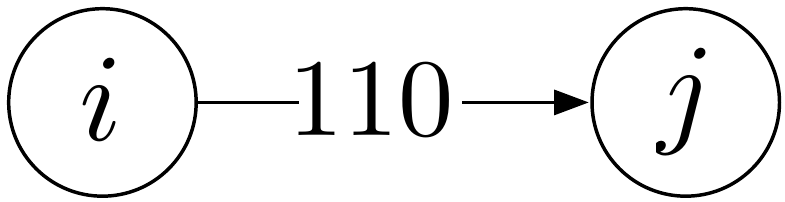}
	 \caption{\label{fig:binary-transform}A example of the $\lambda(\omega)$ mapping.}
\end{figure}

There exists the injective function $\lambda: \Omega \rar \{0,1\}^{\lceil\text{log}_2(|\Omega|)\rceil}$ (a binary encoder) that represents every label in $\Omega$ as a unique binary string of length $\lceil\text{log}_2(|\Omega|)\rceil$. While the minimum bits required to make a one-to-one mapping is $\lceil\text{log}_2(|\Omega|)\rceil$, popular examples of other such one-to-one mappings include the ASCII and Unicode functions that map between human language characters and binary strings. Furthermore, there exist the inverse function $\lambda^{-1}$ that maps a binary string to its original symbolic representation. Note that for labels already represented as unique binary strings, $\lambda$ and $\lambda^{-1}$ are identity functions. Given the semantic network edge diagrammed in Figure \ref{fig:semantic-example}, the $\lambda(\omega)$ mapping is represented in Figure \ref{fig:binary-transform}. Assume that $|\Omega| = 8$ and thus, each $\omega \in \Omega$ requires $3$ bits to encode it.

\begin{figure}[h!]
	\centering
	\includegraphics[width=0.225\textwidth]{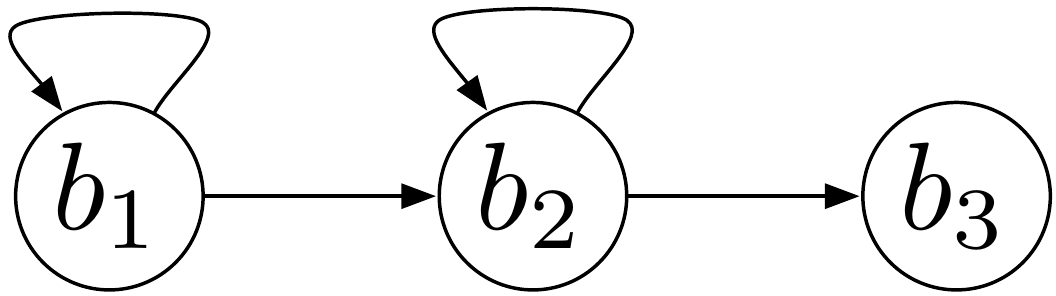}
	 \caption{\label{fig:label-transform}A directed network representation of the edge label $\lambda(\omega) = (1,1,0)$.}
\end{figure}

Next, there exists the injective function $\gamma: \{0,1\}^n \rar \cal{D}$ (a directed network encoder), where $\cal{D}$ is the family of all directed networks and any $D \in \cal{D}$ is denoted $D \subseteq (V^d \times V^d)$. If $B \in \{0,1\}^n$ is the ordered multi-set (or bag) of the $n$-bit string $\lambda(\omega)$, then
\begin{equation*}
	\gamma(B) = \bigcup^{n \leq |B|}_{n=1} 	
	\begin{cases}
		(b_n,b_{n+1}) & \text{if } b_n=0 \wedge n < |B|\\
		(b_n,b_{n+1}) \cup (b_n,b_n) & \text{if } b_n=1 \wedge n < |B| \\
		(b_n,b_n) & \text{if } b_n=1 \wedge n = |B|.
	\end{cases}
\end{equation*}
If $\lambda(\omega) = (1,1,0)$, then $\gamma(\lambda(\omega))$ is represented as diagrammed in Figure \ref{fig:label-transform}. The number of vertices in $D$ with respects to $\gamma$ is $\mathcal{O}(\lceil\text{log}_2(|\Omega|)\rceil)$. The number of directed edges in $D$ with respects to $\gamma$ is $\mathcal{O}(2\lceil\text{log}_2(|\Omega|)\rceil-1)$. 

The function $\gamma$ is information preserving because there also exists the inverse function $\gamma^{-1}$. If $q \in {\{V^d\}}^n$ is the single non-looping path in $D$ that traverses every vertex in $V^d$ (i.e.~the only Hamiltonian path), then
\begin{equation*}
	\gamma^{-1}(D) = \biguplus_{n=1}^{n \leq |q|}
	\begin{cases}
		1 & \text{if } (q_n,q_n) \in D \\
		0 & \text{otherwise}.
	\end{cases}
\end{equation*}
Thus, $\lambda^{-1}(\gamma^{-1}(\gamma(\lambda(\omega)))) = \omega$. From a set of functions that transform a symbolic edge label to a directed network encoding, it is possible to represent an entire semantic network as a a single directed network. In other words, given $\gamma \circ \lambda$, $S \subseteq \la V^{s} \times \cal{D} \times V^{\text{s}} \ra$.

\begin{figure}[h!]
	\centering
	\includegraphics[width=0.4\textwidth]{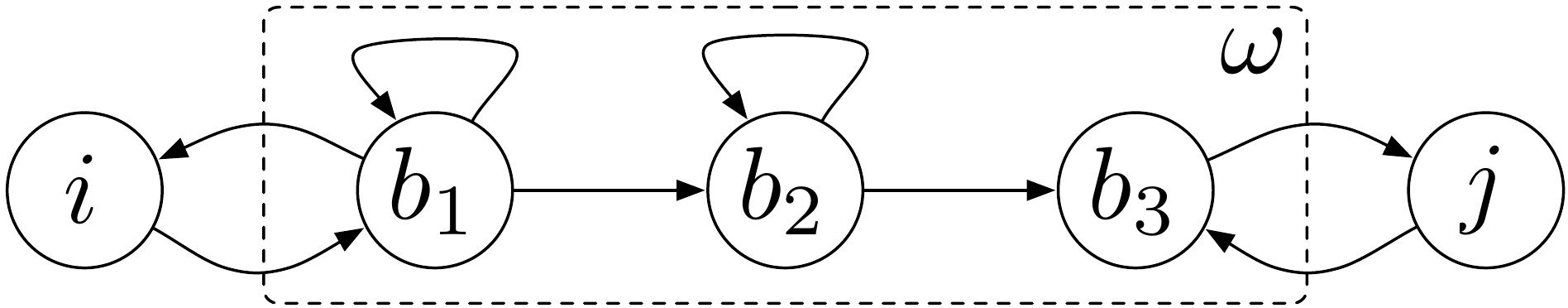}
	 \caption{\label{fig:semantic-nonsemantic}A $D$-encoding of $S$.}
\end{figure}

\begin{proposition}[Semantic-to-Directed Injection]
A semantic network can be modeled as a directed network without loss of information. There exists an injective function $\Theta : \cal{S} \rar \cal{D}$, where $D \in \cal{D}$ is a directed network representation of some $S \in \cal{S}$.
\end{proposition}
\emph{Proof.} If $\Theta : \cal{S} \rar \cal{D}$ denotes an injective function that maps a semantic network to a directed network, then
\begin{equation*}
	\Theta(S) = \bigcup_{\la i,\omega,j \ra \in S} (i,b_1) \cup (b_1,i) \cup \gamma(\lambda(\omega)) \cup (b_n,j) \cup (j,b_n),
\end{equation*}
where any $b$ is a vertex in $\gamma(\lambda(\omega))$ and $n > 1$. With respects to the previous example figures, the $\Theta(S)$ mapping is diagrammed in Figure \ref{fig:semantic-nonsemantic}.

Let $D \subseteq (V^d \times V^d)$ denote the directed network $\Theta(S)$. In $V^d$, every vertex that does not self-loop and has an even degree was originally a vertex in $V^s$. All other vertices in $V^d$ are used to denote the edge labels of $\Omega$. The growth of the number of vertices in $D$ with respects to $\Theta(S)$ is $\mathcal{O}(|V^s| + |S|\lceil\text{log}_2(|\Omega|)\rceil)$. The growth of the number of edges in $D$ with respects to $\Theta(S)$ is $\mathcal{O}(|S|[2\lceil\text{log}_2(|\Omega|)\rceil+3])$.

In order to demonstrate the information-preserving quality of $\Theta$, the inverse function $\Theta^{-1}$ also exists. Let $\Gamma: V^d \rar \mb{N}$ denote the degree of a vertex and let $Q_{i \rar j}$ be the set of paths from vertex $i$ to vertex $j$ in $D$ such that
\begin{equation*}
	Q_{i \rar j} = \bigcup \; (i,b_1,\ldots,b_n,j),
\end{equation*}
where $\frac{|\Gamma(i)|}{2},\frac{|\Gamma(j)|}{2} \in \mb{N}$ (i.e.~$i$ and $j$'s degree is even), $(i,i),(j,j) \notin D$ (i.e.~no self-loops), $(i,b_1),(b_1,i),(b_1,\ldots),(\ldots,b_n),(b_n,j),(j,b_n) \in D$, $i \neq b_1 \neq \ldots \neq b_n$, $j \neq b_1 \neq \ldots \neq b_n$ (i.e.~only $i$ and $j$ can be the same vertex), and no $b$ is in a cycle with another $b$ in the sequence. If
\begin{equation*}
	Q = \bigcup_{i,j \in V^d} Q_{i \rar j},
\end{equation*}
then
\begin{equation*}
	\Theta^{-1}(D) = \bigcup_{q \in Q} \la q_1, \lambda^{-1}(\gamma^{-1}(q_2, \ldots, q_{n-1})), q_n \ra,
\end{equation*}
where $q_1= i$ and $q_n = j$ and thus, the original vertices in $V^s$.

Given $\Theta$ and $\Theta^{-1}$, a unique, one-to-one mapping between a semantic network and a directed network exists such that a semantic network can be modeled as a directed network without loss of information. \qed

There exists another proof of this concept. As demonstrated earlier, a binary string of arbitrary length can be represented as a single chain (i.e.~sequence, path) of vertices, where each vertex represents a bit. In this representation, a self-loop represents a bit with value $1$ and no self-loop represents a bit with value $0$. Because any representation of a semantic network, at the lowest level of computing, is ultimately represented as a sequence of bits, a directed network can  be used to model that sequence.

\section{Mapping a Directed Network to an Undirected Network\label{sec:dir-to-und}}

\begin{figure}[h!]
	\centering
	\includegraphics[width=0.175\textwidth]{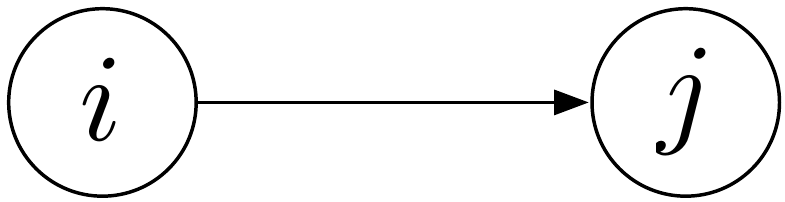}
	 \caption{\label{fig:directed-example}An edge in a directed network.}
\end{figure}

This section presents the injective, information-preserving function $\hat{\Theta} : \cal{D} \rar \cal{U}$ that maps a directed network to an undirected network. A directed network is identified by a set of ordered vertex pairs. For instance, when $D \subseteq (V^d \times V^d)$, $(i,j) \in D$ denotes a directed edge going from $i$ (the source) to $j$ (the sink). A directed edge between $i$ and $j$ is diagrammed in Figure \ref{fig:directed-example}.

An undirected network denoted $U \subseteq \{V^u \times V^u\}$ does not represent edge directionality as elements of $U$ are unordered thus, $\{i,j\}$ states that $i$ and $j$ are connected, but that no particular direction exists. If a directed network is to be represented as an undirected network, then a topological feature in the undirected form must be used to represent edge directionality.

\begin{figure}[h!]
	\centering
	\includegraphics[width=0.25\textwidth]{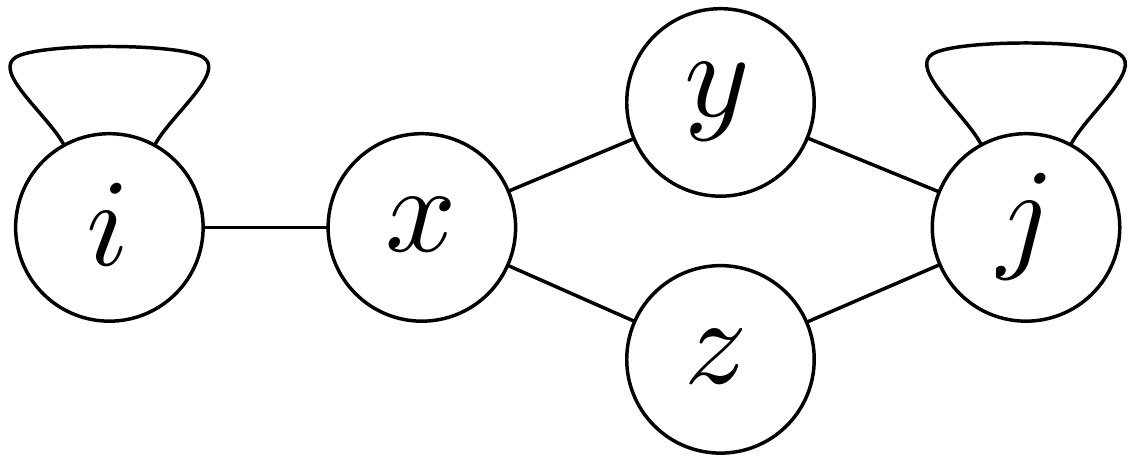}
	 \caption{\label{fig:direction-transform}An undirected network representation of a directed edge.}
\end{figure}

\begin{proposition}[Directed-to-Undirected Injection]
A directed network can be modeled as an undirected network without loss of information. There exists an injective function $\hat{\Theta} : \cal{D} \rar \cal{U}$, where $U \in \cal{U}$ is an undirected network representation of some $D \in \cal{D}$.
\end{proposition}
\emph{Proof.} The function $\hat{\Theta}$ maps each ordered vertex pair in $D$ to a set of unique unordered vertex pairs in $U$. If $R_{i \rar j} = \{i,x\} \cup \{x,y\} \cup \{x,z\} \cup \{y,j\} \cup \{z,j\}$, then
\begin{equation*}
	\hat{\Theta}(D) = \bigcup_{(i,j) \in D} \{i,i\} \cup R_{i \rar j} \cup \{j,j\},
\end{equation*}
where the vertices $x$, $y$, and $z$ are unique for each $(i,j) \in D$. Any vertex with an undirected self-loop in $V^u$ is an original vertex from $V^d$. The vertices $x,y,z \in V^u$ and their respective edges represent the direction of the edge. The vertex $i$ has one edge which denotes the tail of the original directed edge. The vertex $j$ has two edges which denotes the head of the original directed edge. $\hat{\Theta}$ incurs a vertex growth of $\mathcal{O}(|V^d| + 3|D|)$ and an edge growth of $\mathcal{O}(|V^d| + 5|D|)$. The $\hat{\Theta}$ mapping of the directed edge represented in Figure \ref{fig:directed-example} is diagrammed in Figure \ref{fig:direction-transform}.

The function $\hat{\Theta}$ is information preserving because there exists the inverse function $\hat{\Theta}^{-1}$ such that if $q^+: (V \times V) \rar \{0,1\}$ is defined as
\begin{equation*}
	q^+(i,j) = 
	\begin{cases}
		1 & \text{if } \{i,x\},\{x,y\},\{x,z\},\{y,j\},\{z,j\} \in U \\
		0 & \text{otherwise},
	\end{cases}
\end{equation*}
then
\begin{equation*}
	\hat{\Theta}^{-1}(U) = \bigcup_{i,j \in V^u} (i,j) \; : \; \{i,i\},\{j,j\} \in U \; \wedge \; q^+(i,j) = 1.
\end{equation*}
Thus, a directed network can be modeled as an undirected network. \qed

\section{Mapping a Semantic Network to an Undirected Network\label{sec:sem-to-und}}

This section presents the unification of the concepts presented in the two previous sections. In this section, by means of function composition, it is demonstrated that a semantic network can be modeled as an undirected network without loss of information. This means that there exists a one-to-one mapping between a semantic network and some undirected network. In short, given the functions $\Theta$ and $\hat{\Theta}$ presented previously, an undirected network has the same representative or modeling power as a semantic network.

\begin{proposition}[Semantic-to-Undirected Injection]
A semantic network can be modeled as an undirected network without loss of information. There exists an injective function $\hat{\Theta} : \cal{S} \rar \cal{U}$, where $U \in \cal{U}$ is an undirected network representation of some $S \in \cal{S}$.
\end{proposition}
\emph{Proof.} Recall the injective functions $\Theta : \cal{S} \rar \cal{D}$ and $\hat{\Theta} : \cal{D} \rar \cal{U}$. Through function composition, there exists the function $\Upsilon : \cal{S} \rar \cal{U}$ with the rule
\begin{equation*}
	\Upsilon(S) = \hat{\Theta}(\Theta(S)).
\end{equation*}
$\Upsilon$ incurs a vertex growth of 
\begin{equation*}
\mathcal{O}([|V^s| + 7|S|\lceil\text{log}_2(|\Omega|)\rceil + 9|S|)
\end{equation*}
and an edge growth of
\begin{equation*}
\mathcal{O}([|V^s| + 11|S|\lceil\text{log}_2(|\Omega|)\rceil + 15|S|). 
\end{equation*}

Finally, there also exists the inverse function $\Upsilon^{-1}$, where
\begin{equation*}
	\Upsilon^{-1}(U) = \Theta^{-1}(\hat{\Theta}^{-1}(U)).
\end{equation*}
Thus, a semantic network can be modeled as an undirected network. \qed

\begin{figure}[h!]
	\centering
	\includegraphics[width=0.4975\textwidth]{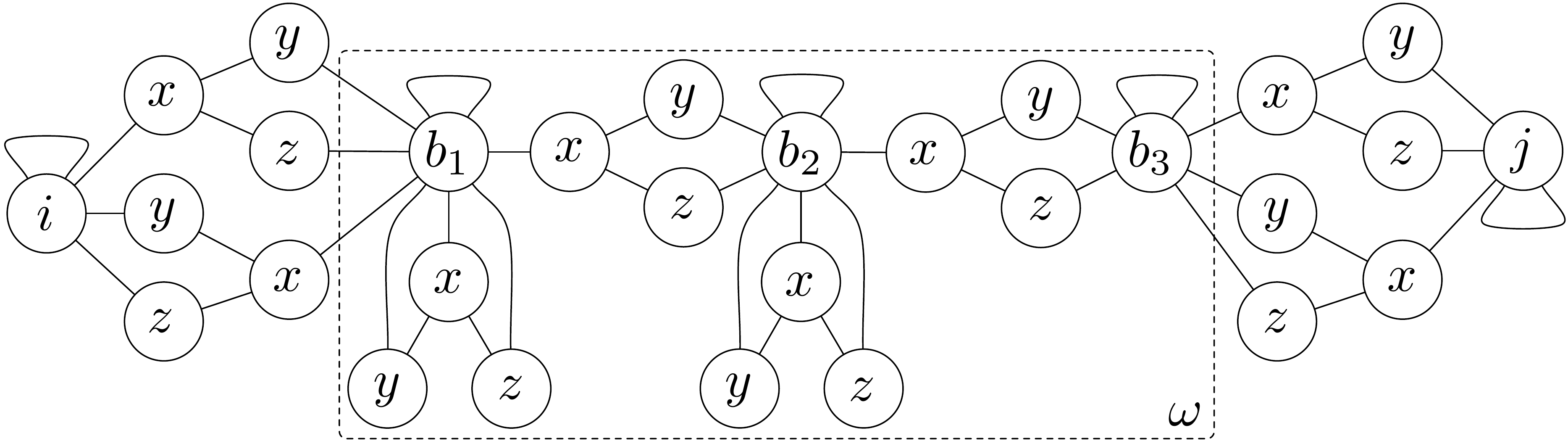}
	 \caption{\label{fig:semantic-undirected}An undirected network representation of a semantic network triple.}
\end{figure}

Given the example semantic network triple diagrammed in Figure \ref{fig:semantic-example}, where $S = \la i,\omega,j \ra$ and $\lambda(\omega) = (1,1,0)$, the undirected network representation given by $\Upsilon(S)$ is diagrammed in Figure \ref{fig:semantic-undirected}. Note that each $x$, $y$, and $z$ is a unique vertex even though they are not notated as such.

It is interesting to note the various types of self-loops in the undirected network representation in Figure \ref{fig:semantic-undirected}. There are the undirected self-loops as demonstrated by the edges $\{i,i\}$, $\{b_n,b_n\}$, and $\{j,j\}$. Next, there are the directed self-loops as demonstrated by the $b_1$ and $b_2$ sub-networks which include their respective $x,y,z$ vertices. Finally, if $i = j$, there also exists the semantic self-loop.

There exists another method to map a semantic network to an undirected network. As discussed previously,  a directed network can represent a binary string and any semantic network representation, computationally, is ultimately represented as a series of bits. Therefore, it is possible to represent a semantic network as a directed network binary string. Given $\hat{\Theta}$, it is possible to represent that directed network binary string as an undirected network.

\section{Conclusion}

This article defined the injective function $\Upsilon: \cal{S} \rar \cal{U}$. This function demonstrates that a semantic network has a one-to-one mapping with some undirected network. In this model, because an edge in an undirected network is neither labeled nor directed, both the semantic network edge labels and the directionality of edges are represented as topological features of the undirected network. While representing a semantic network as an undirected network is perhaps an inefficient use of resources, it is theoretically possible.

\end{document}